# Spectroscopy of Candidate Young Globular Clusters in NGC 1275

Stephen E. Zepf[1]

Department of Astronomy, University of California, Berkeley, CA 94720
e-mail: zepf@astron.berkeley.edu

Dave Carter

Royal Greenwich Observatory, Madingley Road, Cambridge CB3 0EZ, UK
e-mail: dxc@mail.ast.cam.ac.uk

Ray M. Sharples

Department of Physics, University of Durham, South Road, Durham, DH1 3LE, UK
e-mail: r.m.sharples@durham.ac.uk

and

Keith M. Ashman

Department of Physics and Astronomy, University of Kansas, Lawrence, KS 66045
e-mail: ashman@kusmos.phsx.ukans.edu

astro-ph/9503050    12 Mar 1995

---

[1] Hubble Fellow

# ABSTRACT

We present spectra of the brightest member of the population of compact blue objects discovered in the peculiar galaxy NGC 1275 by Holtzman et al. using Hubble Space Telescope images. These spectra show strong Balmer absorption lines like those observed in A-type stars, as expected if the object is a young globular cluster. The age estimated from the strength of the Balmer lines is about 0.5 Gyr, although ages ranging from 0.1 Gyr to 0.9 Gyr can not be confidently excluded given current models of stellar populations. If these estimated ages are adopted for the young cluster population of NGC 1275 as a whole, the fading predicted by stellar populations models gives a luminosity function which is consistent with that of the Galactic globular cluster system convolved with the observational selection function for the NGC 1275 system. We also use the equivalent widths of the Mg $b$ and Fe 5270 features to constrain the metallicity of the young cluster. Combining these absorption-line widths with the age estimates from the Balmer lines and stellar population models, we find a metallicity of roughly solar based on the Mg $b$ index and somewhat higher for the Fe 5270 index. The radial velocity of the absorption lines of the cluster spectrum is offset from the emission lines of the galaxy spectrum at the same position by $-130$ km s$^{-1}$, providing further evidence for the identification of the object as a globular cluster, and opening up the future possibility of studying the kinematics of young cluster systems. The discovery of objects with the characteristics of young globular clusters in NGC 1275, which shows evidence of a recent interaction or merger, supports the hypothesis that galaxy interactions and mergers are favorable sites for the formation of globular clusters.

*Subject headings:* galaxies: individual (NGC 1275) — galaxies: interactions — galaxies: formation —galaxies: star clusters

## 1.  Introduction

A large number of bright, blue compact sources with sizes, colors, and luminosities consistent with those of a population of young globular clusters were discovered in the peculiar galaxy NGC 1275 by Holtzman et al. (1992) using HST images. This observation and the discovery of similar objects in the prototypical merger NGC 7252 (Whitmore et al. 1993) suggest that galaxy interactions and mergers are favorable environments for the formation of globular clusters. These results were predicted by Ashman & Zepf (1992) based on the hypothesis that elliptical galaxies are made from the merger of spiral galaxies. If these bright, blue compact sources are confirmed to be globular clusters, then the greater frequency of globular clusters around ellipticals relative to spirals (e.g. Harris 1991) may be accounted for naturally in the context of a merger origin for elliptical galaxies. Moreover, such a discovery would open up the possibility of studying the process of globular cluster formation in its early phases.

Spectroscopy of the candidate young globulars identified in HST images is critical for establishing their nature. Spectroscopy directly addresses the question of whether the light is from stars, and is a powerful tool for studying the stellar content of objects shown to be star clusters. In particular, the strength of the Balmer lines is a sensitive and reddening-indepedent discriminant of the age of a stellar population. Determining the age is important because it plays a major role in the calculation of the mass-to-light ratio using models of stellar populations. Spectroscopy also provides constraints on the metallicity of the young clusters, which has significant implications for theories of globular cluster formation. Moreover, a radial velocity can be determined from the spectra, which allows for the removal of any contamination from objects not associated with NGC 1275.

In this letter, we present the first results of our program to obtain spectra of candidate young globular clusters identified in high resolution images. In Section 2, we present observations of the young cluster system in NGC 1275. In Section 3 we describe the age, velocity, and metallicity determined for the brightest of the candidate clusters. The implications of these results and directions for future work are discussed in Section 4.

## 2.  Observations

Spectra of the objects around NGC 1275 identified as candidate young globular clusters were obtained on 18 and 20 Oct 1993 with the 4.2m William Herschel Telescope. The Low Dispersion Survey Spectrograph (LDSS-2, Allington-Smith et al. 1994) was used in "high" dispersion mode, giving useful spectral coverage from 3700 Å to 5500 Å at a resolution of 4.5 Å (FWHM). The spatial scale of the instrument is 0.55 arcsec/pixel. This letter focusses on the central slitlet of each multi-slit mask, which was centered on the brightest of the candidate young clusters. The slitlets were oriented at a position angle of 0° on the first night, and 30° on the second night of observations. An overlay of the slit positions on an HST image of the center of NGC 1275 is shown in Figure 1 [Plate 00].

Our observations consist of a series of 30 minute exposures totalling 5 hours at the two position angles. CCD data reduction and wavelength calibration were performed using standard IRAF packages. From the two dimensional spectra, apertures were extracted at positions centered on the candidate young clusters. These are readily visible as enhancements along the slit profile, particularly in frames with good seeing. Apertures offset several arcseconds in each direction were then extracted to provide the basis for the subtraction of the bright galaxy background. The spectra were flux calibrated using a narrow slit observation of the spectrophotometric standard G191B2B (Massey et al. 1988) and dereddened assuming E($B - V$) = 0.17 in the direction of NGC 1275 (Burstein & Heiles 1984) and no internal reddening. This procedure gives a precise relative flux calibration, which is sufficient for determining absorption-line indices.

In Figure 2, we plot the extracted spectra at the position of the brightest cluster (N1275-H1) and galaxy background apertures. These spectra are the sum of the three exposures at each position angle with the best image quality, which is estimated to be $0.7'' - 0.8''$. All subsequent analysis is based on this subset. This figure shows that the spectra from the apertures centered on N1275-H1 have much stronger Balmer absorption lines relative to the continuum and emission lines than either of the two galaxy apertures, as expected if N1275-H1 is a young star cluster.

A quantitative study of the object spectrum requires accurate background subtraction, since the galaxy contributes about half of the light within a one pixel aperture centered on the brightest object. Predicting the overall level of the contribution from the galaxy background is straightforward. However, determining the detailed galaxy spectrum at this point is more complicated because the spectrum of the central region of NGC 1275 varies with position. The variation of the strength and velocity of the emission lines is particularly problematic. We estimate the galaxy background by combining the galaxy spectra from either side of the object, subject to the constraints that the galaxy flux level at the object aperture is correct, and that the equivalent widths of the [O II] and [O III] emission lines of the resulting object spectrum are zero.

The background-subtracted spectra of N1275-H1 are presented in Figure 3. The top two panels are the spectra for the slitlets at position angles of 0° and 30° respectively. The bottom panel is the spectrum of the A star SAO 76504 obtained immediately after the observations of NGC 1275 using the same instrumental set-up. The spectrum of N1275-H1 clearly appears to be that of a young star cluster in which the light is dominated by A stars. The reality of the strong Balmer absorption lines in these spectra of N1275-H1 and the reliability of the background subtraction procedure is supported by a number of tests. Firstly, the results for the spectra taken at different position angles are in good agreement, even though the background is taken from different regions of the galaxy in the two cases. Secondly, the object spectra which result from the background subtraction have line ratios and an overall shape consistent with those expected from a young stellar population. Finally, the strong Balmer absorption is visible in the aperture on N1275-H1 before the background subtraction (Figure 2).

### 3. Analysis

The identification of N1275-H1 as a young star cluster is strongly supported by the spectra presented in Figure 3, which show that this object has strong, broad Balmer absorption lines, like those observed in A and early F type stars. A more quantitative analysis of the age and metallicity of N1275-H1 can be made by comparing the strength of various absorption lines to models of stellar populations. In principle, the age and metallicity of young star clusters can be constrained by determining the equivalent widths of the Balmer absorption lines, which rise to a peak at about 0.5 Gyr and then decline for older ages (e.g. Bica & Alloin 1986, Schweizer & Seitzer 1993), and the strength of the metallic absorption lines, which generally increases monotonically with age and metallicity (e.g. Fritze-v. Alvensleben & Burkert 1995). Constraining the age through the strength of the Balmer lines is particularly valuable, since age is the most important factor in estimates of the mass-to-light ratio, and thus of comparisons of the masses of the young globular clusters to the Galactic globular cluster system.

In order to compare the absorption lines in our N1275-H1 spectrum to models and other observations, we use a number of previously defined absorption-line indices. For metallic features, we adopt the definition of the Mg $b$ and Fe 5270 indices of Faber et al. (1985). Because the Balmer line indices in this system are narrower than the broad absorption observed in young stellar populations, we adopt the broader definitions of Brodie and Hanes (1986) for the Balmer lines, with a minor shift of 5 Å in the bandpass of the H$\delta$ index. The equivalent widths in these indices for the spectra shown in Figure 3 are listed in Table 1. We estimate the uncertainty in these equivalent widths by the difference between the independent measurements at different position angles, which is typically 2-3 Å for the Balmer lines and 0.5-1.0 Å for the metal lines.

The strong Balmer absorption typical of A stars is clearly evident in the equivalent widths given in Table 1. Ideally, we would simply compare these widths to a stellar populations model and determine the age. However, the commonly used Bruzual & Charlot (1993) models are unable to produce Balmer absorption-line widths as strong as those observed at any age. A similar effect was noted for two young globular clusters in NGC 7252 by Schweizer & Seitzer (1993). The discrepancy remains when the data are smoothed to match the spectral resolution of the models. Most variations of the model IMF are also ineffective at mitigating this difference, because the continuum is dominated by the same stars which produce the strong Balmer line absorption. The largest Balmer line equivalent widths predicted by the Bruzual & Charlot model with a standard IMF occur at an age of 0.5 Gyr, at which the model underpredicts the observed Balmer line widths by about 30%.

An alternative approach is to compare the absorption-line indices of N1275-H1 to those of the integrated spectra of globular clusters for which age estimates are available from color-magnitude diagrams. Using the globular cluster sample of Bica & Alloin (1986), we find that the Balmer line equivalent widths of N1275-H1 are best matched by LMC clusters with ages of 0.5 Gyr. At these ages, the LMC clusters have Balmer line equivalent widths which are very similar to those of N1275-H1. Although this comparison gives a best fit age of about 0.5 Gyr, ages in the range 0.1 - 0.9 Gyr cannot be ruled out because of the small sample size and observational scatter,

The age estimate for N1275-H1 can be combined with stellar population models to predict the luminosity function of the young cluster system in NGC 1275 at an age comparable to that of the Galactic globular cluster system. In the $V$ band, the Bruzual & Charlot models predict that a 0.5 Gyr old system will become 3.7 mag fainter at an age of 15 Gyr. The corresponding fading for 0.1 Gyr and 0.9 Gyr old systems is 5.3 and 3.2 mag respectively. These roughly bracket the amount of fading to required to bring the observed luminosity function of the young globular cluster system of NGC 1275 into agreement with the luminosity function of the Galactic globular cluster system. However, deeper HST observations are required to test whether the turnover observed in the luminosity function of the young clusters is real or a result of incompleteness. Moreover, the fading calculation assumes a uniform age in the young cluster population, which has not been demonstrated conclusively (e.g. Faber 1993).

In principle, optical colors provide an independent constraint on the age of the young clusters. However, the reddening dependence of the colors makes it difficult to apply this technique to the young clusters of NGC 1275, which lies at relatively low Galactic latitude, and may also have significant internal reddening. With this caveat, we note that the optical colors given by Holtzman et al. (1992) and Richer et al. (1993) are best fit by an age at the young end of the range indicated by the Balmer absorption-line widths. This difference may be reconciled either by decreasing the assumed reddening or by adopting the youngest age consistent with the Balmer absorption lines. In either case, the predicted luminosity function for the young cluster system of NGC 1275 at an age of 15 Gyr tends toward the fainter end of the range given above.

The Mg $b$ and Fe 5270 absorption-line indices can be used to estimate the metallicity of the young cluster. The metallicity is important for understanding the origin of the gas out of which the young cluster system formed and for testing theories of globular cluster formation. Although the indices are dependent on both age and metallicity, we can place limits on the metallicity by considering the indices over the range of ages determined above. For the Mg $b$ index, the values in Table 1 bracket the values predicted by the solar metallicity Bruzual & Charlot model for the age range from 0.1 to 0.9 Gyr. However, the Fe 5270 index is significantly higher than predicted by these models over the given age range. The models of Fritze-v. Alvensleben & Burkert (1995) suggest that the Fe 5270 index indicates an abundance of at least several times solar. Given the observational and modelling uncertainties, we conclude that the metallicity of N1275-H1 is consistent with solar, and wait for better models and data before constraining the metallicity or abundance ratios in more detail.

By comparing the redshift of the Balmer absorption lines in NGC 1275-H1 to the redshift found for the emission lines in the same aperture, we can determine the velocity of N 1275-H1 relative to the emission line gas of NGC 1275. For the PA = 0° spectrum, we find that N1275-H1 is at a velocity of $-140 \pm 25$ km s$^{-1}$ with respect to the emission lines, and for the PA = 30° spectrum we find an offset of $-120 \pm 15$ km s$^{-1}$. This velocity clearly shows that N1275-H1 is a distinct object physically associated with the galaxy.

## 4. Discussion

The spectrum of N1275-H1 shows that its optical light is dominated by A-type stars, as expected if it is a young globular cluster. Combined with the similar result of Schweizer & Seitzer (1993) for two of the candidate 3 young globulars in NGC 7252, this provides strong observational evidence for the formation of globular clusters in galaxy interactions and mergers. If future observations show that the globular cluster formation process is as efficient as rough initial estimates suggest (Zepf & Ashman 1993), then one of the main objections to the formation of ellipticals through the merger of spirals will have been removed. An episodic formation history for elliptical galaxies is also consistent with recent observations of the color distributions of their globular clusters (e.g. Zepf, Ashman, & Geisler 1995).

The study of young globular cluster systems also opens an observational window on the process of globular cluster formation. As an example, both N1275-H1 and the two young clusters in NGC 7252 have metallicities which are estimated to be roughly solar based on absorption-line indices. These observations clearly suggest that globular clusters can form over a wide range of metallicities. It may also soon be possible to obtain velocities for a number of young clusters in NGC 1275 and other galaxy mergers and thereby constrain the kinematics of young globular cluster systems.

We acknowledge the assistance of Dr. N. Tanvir in obtaining archival HST images of NGC 1275. S.E.Z. acknowledges support from NASA through grant number HF-1055.01-93A awarded by the Space Telescope Science Institute, which is operated by the Association of Universities for Research in Astronomy, Inc., for NASA under contract NAS5-26555. K.M.A. acknowledges support from a Fullam/Dudley Award and a Dunham Grant from the Fund for Astrophysical Research. The spectroscopy reported here was obtained with the William Herschel Telescope operated on the island of La Palma by the Royal Greenwich Observatory in the Spanish Observatorio del Roque de los Muchachos of the Instituto de Astrofísica de las Canarias.

# REFERENCES

Allington-Smith, J. et al. 1994, PASP, 106, 983

Ashman, K.M., & Zepf, S.E. 1992, ApJ, 384, 50

Bica, E., & Alloin, D. 1986, A&A, 162, 21

Brodie, J.P. & Hanes, D.A. 1986, ApJ, 300, 258

Bruzual, A.G., & Charlot, S. 1993, ApJ, 405, 538

Burstein, D., & Heiles, C.. 1984, ApJS, 54, 33

Cardelli, J., Clayton, & Mathis 1989, ApJ, 345, 245

Faber, S.M. 1993, in The Globular Cluster-Galaxy Connection, eds. G.H. Smith & J.P. Brodie (San Francisco: ASP), 601

Faber, S.M., Friel, E., Burstein, D., & Gaskell, C.M. 1985, ApJS, 57, 711

Fritze-v. Alvensleben, U., & Burkert, A. 1995, A&A, in press

Harris, W.E. 1991, ARAA, 29, 543

Holtzman, J.A. et al. 1992, AJ, 103, 691

Massey, P., Strobel, K., Barnes, J.V., & Anderson, E. 1988, ApJ, 328, 315

Richer, H.B., Crabtree, D.R., Fabian, A.C., & Lin, D.N.C. 1993, AJ, 105, 877

Schweizer, F., & Seitzer, P. 1993 ApJ, 417, L29

Whitmore, B.C., Schweizer, F., Leitherer, C., Borne, K., & Robert, C. 1993, AJ, 106, 1354

Zepf, S.E., & Ashman, K.M. 1993 MNRAS, 264, 611

Zepf, S.E., Ashman, K.M., & Geisler, D. 1995, ApJ, in press

**TABLE 1: Absorption-Line Widths for NGC 1275-H1.**

| Object | EW (H$\beta$) Å | EW (H$\gamma$) Å | EW (H$\delta$) Å | EW (Mg $b$) Å | EW (Fe 5270) Å |
|---|---|---|---|---|---|
| N1275-H1 PA=0° | 14.7 | 10.7 | 13.3 | 0.6 | 2.6 |
| N1275-H1 PA=30° | 11.5 | 10.0 | 8.3 | 1.3 | 1.9 |
| SAO 76504 | 12.7 | 9.7 | 8.4 | 0.6 | 1.2 |

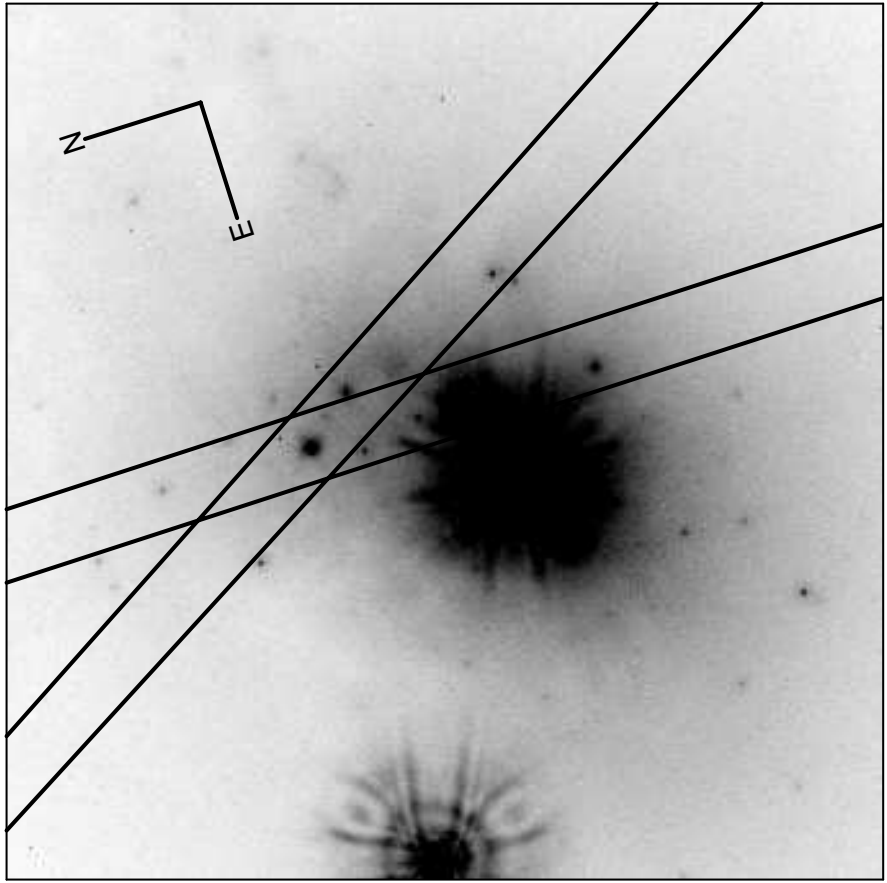

Fig. 1.— The positions of the slits overlayed on a WFPC-1 image of NGC 1275 in the F555W filter. N1275-H1 is the object at the intersection of the two slits. The image is 15″ on a side. No deconvolution was performed on the image.

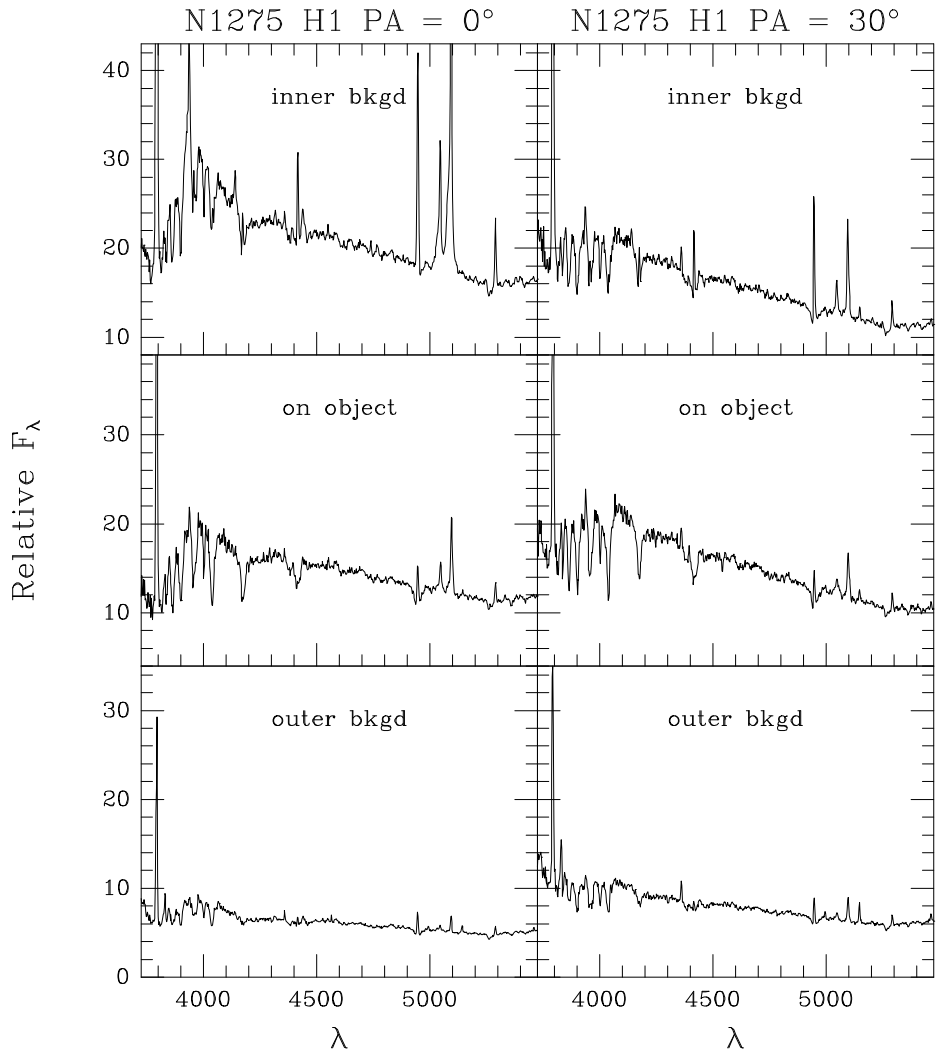

Fig. 2 - A plot of the spectra within the various apertures used to determine the spectrum of
N1275-H1. Those corresponding to a slit PA = 0° are on the left and those for PA = 30° are on
the right. Both position angles show stronger absorption in the Balmer lines for the spectrum
from the aperture at the object position compared to either background aperture.

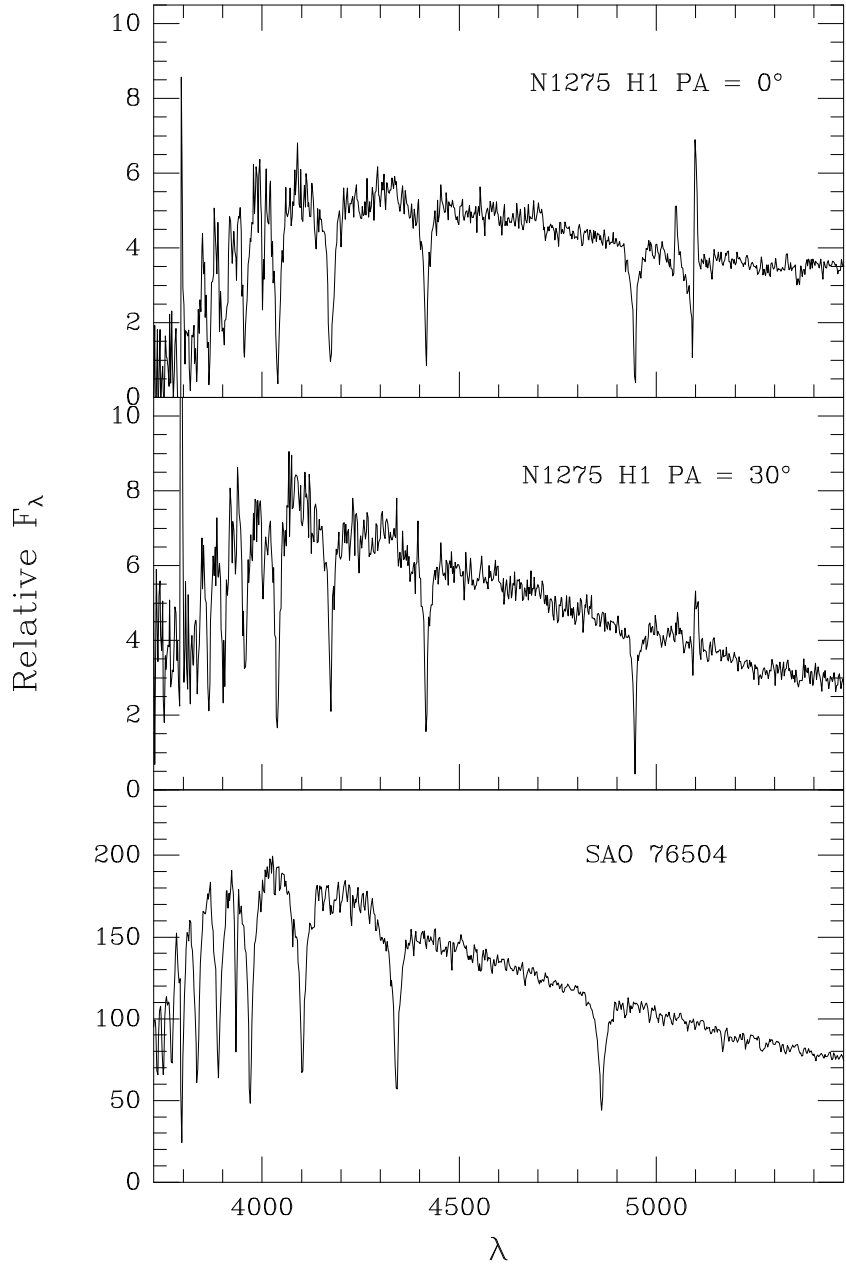

Fig. 3 - The spectrum of N1275-H1 as determined from slitlets at the given position angles is plotted. The bottom panel is the spectrum of an A-star obtained with the same instrument during one of the nights the object spectra were obtained. The N1275-H1 spectra are strikingly similar to the spectrum of the A star, demonstrating that N1275-H1 is a young star cluster.